\documentclass[twocolumn,twoside,slac_two]{revtex4}
\usepackage{graphicx}
\usepackage{fancyhdr}
\usepackage{amsmath}
\begin{document}

%Title of paper
\title{\boldmath Inclusive $|V_{cb}|$ and global fit}

% Repeat the \author .. \affiliation  etc. as needed
%
% \affiliation command applies to all authors since the last
% \affiliation command. The \affiliation command should follow the
% other information

\author{C. Schwanda}
\affiliation{Institute of High Energy Physics, Vienna, Austria}

\begin{abstract}
A reliable determination of the Cabibbo-Kobayashi-Maskawa matrix
element $|V_{cb}|$ is mandatory for precision flavor physics and for
the search for $CP$~violating phases from new, heavy particles. In
this article, we review the theory of the determination of $|V_{cb}|$
from inclusive semileptonic $B$~decays. We discuss the available
measurements of the semileptonic $B$~branching fraction and other
inclusive observables in $B$~decays relevant to the determination of
$|V_{cb}|$. Finally, we perform a global fit to extract $|V_{cb}|$ and
the $b$-quark mass $m_b$.
\end{abstract}

%\maketitle must follow title, authors, abstract
\maketitle

\thispagestyle{fancy}

% body of paper here - Use proper section commands
% References should be done using the \cite, \ref, and \label commands
% Put \label in argument of \section for cross-referencing
%\section{\label{}}
\section{Introduction}

The magnitude of the Cabibbo-Kobayashi-Maskawa (CKM) matrix element
$|V_{cb}|$~\cite{Kobayashi:1973fv} can be determined from semileptonic
$B$~decays to charmed final states. At the $B$~factory experiments
Belle~\cite{unknown:2000cg} and BaBar~\cite{Aubert:2001tu},
measurements of semileptonic decays to $D$ or $D^*$ mesons
(``exclusive measurements'') and determinations of $|V_{cb}|$ using
all semileptonic final states in a given region of phase space
(``inclusive measurements'') have been performed. In this article, we
will review the latter approach. As experimental and theoretical
systematics are different, consistency between exclusive and inclusive
determinations of $|V_{cb}|$ is crucial for our understanding.

After reviewing the theory of the determination of $|V_{cb}|$ from
inclusive $B$~decays, we will discuss the measurements (mainly at the
B~factories) of the semileptonic branching fraction and other
inclusive observables in $B\to X_c\ell\nu$~decays that allow to
determine the non-perturbative parameters that appear in the
calculation of the semileptonic width. Finally, we perform fits to
different data sets to determine $|V_{cb}|$ and the $b$-quark
mass~$m_b$.

\section{\boldmath Calculations of the semileptonic $B$~decay width}
\label{sect:2}

The semileptonic decay width $\Gamma(B\to X_c\ell\nu)$ is proportional
to $|V_{cb}|^2$. Measurements of the semileptonic $B$~branching
fraction thus allow to determine $|V_{cb}|$, provided that the width
can be calculated reliably. Challenges are non-perturbative QCD
contributions and experimental selections applied to the data. {\it
  E.g.}, semileptonic decays can typically only be measured above a
certain minimum lepton energy threshold.

These calculations are performed in the frameworks of the Heavy Quark
Effective Theory (HQET) and the Operator Production Expansion
(OPE)~\cite{Bauer:2004ve,Benson:2003kp}. The result is an expansion in
inverse powers of the $b$-quark mass, the leading order corresponding
to the result obtained assuming unconfined quarks. A problem for the
practical use of these formulae is that new, so-called heavy quark (HQ)
parameter appear at each order in $1/m_b$. These non-perturbative
quantities encode the soft QCD physics and cannot be calculated from
perturbation theory.

At present, there are two independent calculations of the semileptonic
width performed up to the third order in
$1/m_b$~\cite{Bauer:2004ve,Benson:2003kp}. Refering to the $b$-quark
mass definition used, the two schemes are called 1S and kinetic. {\it
  E.g.}, the result in the kinetic scheme reads~\cite{Benson:2003kp},
\begin{multline}
  \Gamma_\mathrm{s.l.}=\frac{G^2_Fm_b^5(\mu)}{192\pi^3}|V_{cb}|^2(1+A_\mathrm{ew})A^\mathrm{pert}(r,\mu)\\
  \Big[z_0(r)\left(1-\frac{\mu^2_\pi(\mu)-\mu^2_G(\mu)+\frac{\rho^3_D(\mu)+\rho^3_{LS}(\mu)}{m_b(\mu)}}{2m_b^2(\mu)}\right)\\
    -2(1-r)^4\frac{\mu^2_G(\mu)-\frac{\rho^3_D(\mu)+\rho^3_{LS}(\mu)}{m_b(\mu)}}{m_b^2(\mu)}\\
    +d(r)\frac{\rho^3_D(\mu)}{m^3_b(\mu)}+\dots\Big]~, \label{eq:1}
\end{multline}
where $G_F$ is the Fermi constant, $\mu$ is the renormalization scale,
$r=m^2_c(\mu)/m^2_b(\mu)$, and $z_0(r)$ and $d(r)$ are phase space
factors~\cite{Benson:2003kp}. Both the electroweak and the
perturbative corrections, $A_\mathrm{ew}$ and $A^\mathrm{pert}(r,\mu)$,
are well known~\cite{Benson:2003kp}. The remaining quantities are the
above-mentioned HQ parameters shown in Table~\ref{tab:1} together with
their counterparts in the 1S scheme.
\begin{table}
  \caption{Heavy quark parameters in the expressions derived in the
  kinetic and 1S schemes.} \label{tab:1}
  \begin{center}
    {\small \begin{tabular}{c|c|c}
      \hline \hline
      & Kinetic scheme & 1S scheme\\
      \hline
      $\mathcal{O}(1)$ & $m^\mathrm{kin}_b$, $m^\mathrm{kin}_c$ &
      $m^\mathrm{1S}_b$\\
      $\mathcal{O}(1/m^2_b)$ & $\mu^2_\pi$, $\mu^2_G$ & $\lambda_1$,
      $\lambda_2$\\
      $\mathcal{O}(1/m^3_b)$ & $\rho^3_D$, $\rho^3_{LS}$ & $\rho_1$,
      $\tau_1$, $\tau_2$, $\tau_3$\\
      \hline \hline
    \end{tabular} }
  \end{center}
\end{table}

Although these HQ parameters cannot be determined theoretically, they
can be obtained from experiment: Other inclusive observables in
$B$~decays, such as the moments of the lepton energy and the
$X_c$~mass distributions in $B\to X_c\ell\nu$~decays, and the
moments of the photon energy spectrum in $B\to X_s\gamma$~decays, have
a similar expansion in inverse powers of $m_b$ with the \emph{same}
HQ parameters. By measuring these quantities in the experiment, one
can determine the parameters in Table~\ref{tab:1} and substitute them
into Eq.~\ref{eq:1} to obtain $|V_{cb}|$ with a precision of
$1-2\%$. This type of analysis is often refered to as \emph{global
  fit}.

On the theory side, this approach requires calculations of the $B\to
X_c\ell\nu$ and $B\to X_s\gamma$ moments with the same precision in
$1/m_b$ as in Eq.~\ref{eq:1}. In the kinetic scheme, calculations of
the lepton energy and hadronic mass spectra up to
$\mathcal{O}(1/m^3_b)$ for different truncations in the lepton energy
are available~\cite{Gambino:2004qm}. Also, the moments of the photon
energy spectrum in $B\to X_s\gamma$~decays have been
calculated~\cite{Benson:2004sg}. Note that the photon energy spectrum
cannot be described by the OPE alone but some modeling of additional
non-perturbative contributions is necessary. In
Ref.~\cite{Benson:2004sg} these are refered to as bias
corrections. Calculations up to the same order in $1/m_b$ are also
available in the 1S scheme~\cite{Bauer:2004ve}.

\section{\boldmath Measurements of $B\to X_c\ell\nu$~decay
  distributions}

Measurements of the semileptonic $B$~branching fraction and inclusive
observables in $B\to X_c\ell\nu$~decays have been obtained by the
BaBar~\cite{Aubert:2004td,Aubert:2004tea,Aubert:2007yaa},
Belle~\cite{Urquijo:2006wd,Schwanda:2006nf}, CDF~\cite{Acosta:2005qh},
CLEO~\cite{Csorna:2004kp} and DELPHI~\cite{Abdallah:2005cx}
experiments. The photon energy spectrum in $B\to X_s\gamma$~decays has
been studied by BaBar~\cite{Aubert:2006gg,Aubert:2005cua},
Belle~\cite{Schwanda:2008kw,Abe:2008sxa} and
CLEO~\cite{Chen:2001fja}. In this section, we will briefly review the
new or updated measurements of $B\to X_c\ell\nu$~decays.

\begin{figure}
  \begin{center}
    \includegraphics[width=0.98\columnwidth]{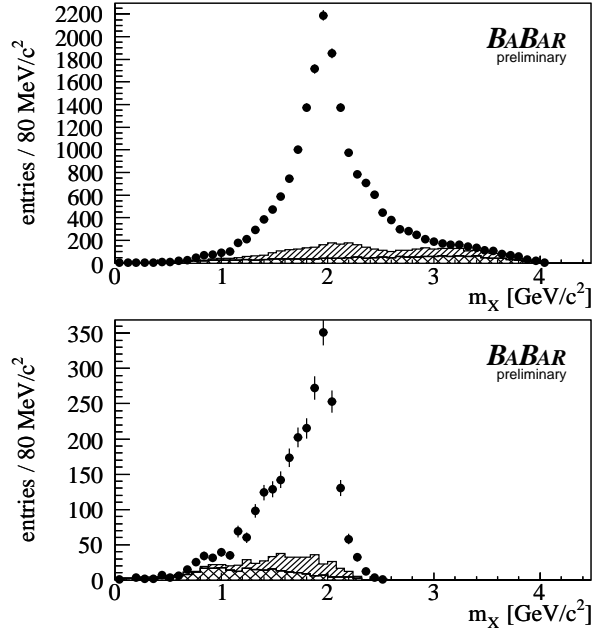}
  \end{center}
  \caption{Hadronic mass spectra in BaBar's recent
    analysis~\cite{Aubert:2007yaa} for minimum c.m.\ lepton momenta of
    0.8~GeV/$c$ (top) and 1.9~GeV/$c$ (bottom). The residual
    backgrounds are shown by the histograms.} \label{fig:1}
\end{figure}

BaBar has updated their previous measurement of the hadronic mass moments
$\langle M^{2n}_X\rangle$~\cite{Aubert:2004tea} and obtained preliminary
results based on a dataset of 210~fb$^{-1}$ taken at the
$\Upsilon(4S)$~resonance~\cite{Aubert:2007yaa}. In the updated
analysis, the hadronic decay of one $B$~meson in $\Upsilon(4S)\to
B\bar B$ is fully reconstructed ($B_\mathrm{tag}$) and the
semileptonic decay of the second $B$ is infered from the presence of
an identified lepton ($e$ or $\mu$) within the remaining particles in
the event ($B_\mathrm{sig}$). This so-called full reconstruction tag
allows to significantly reduce combinatorial backgrounds and select
semileptonic decays with a purity of about 80\%. Particles used
neither for the reconstruction of $B_\mathrm{tag}$ nor for the charged
lepton are considered to belong to the $X_c$~system, and the hadronic
mass spectrum $M_X$ is calculated using some kinematic constraints
(Fig.~\ref{fig:1}).

From this spectrum, BaBar calculates the hadronic mass moments
$\langle M^n_X\rangle$, $n=1,\dots,6$ for minimum lepton momenta in
the center-of-mass (c.m.) frame ranging from 0.8 to 1.9~GeV/$c$. These
moments are however distorted by acceptance and finite resolution
effects and an event-by-event correction is derived from Monte Carlo
(MC) simulated events. These corrections are linear functions with
coefficients depending on the lepton momentum, the multiplicity of the
$X_c$~system and $E_\mathrm{miss}-c|\vec p_\mathrm{miss}|$, where
$E_\mathrm{miss}$ and $\vec p_\mathrm{miss}$ are the missing energy
and 3-momentum in the event. Note that this analysis
measures also mixed mass and c.m.\ energy moments $\langle N^{2n}_X\rangle$,
$n=1,2,3$, with $N_X=M^2_Xc^4-2\tilde\Lambda E_X+\tilde\Lambda^2$ and
$\tilde\Lambda=0.65$~GeV in addition to ordinary hadronic mass
moments. These mixed moments are expected to better constrain some HQ
parameters though they are not used in current global fit analyses
yet.

\begin{figure}
  \begin{center}
    \includegraphics[width=0.9\columnwidth]{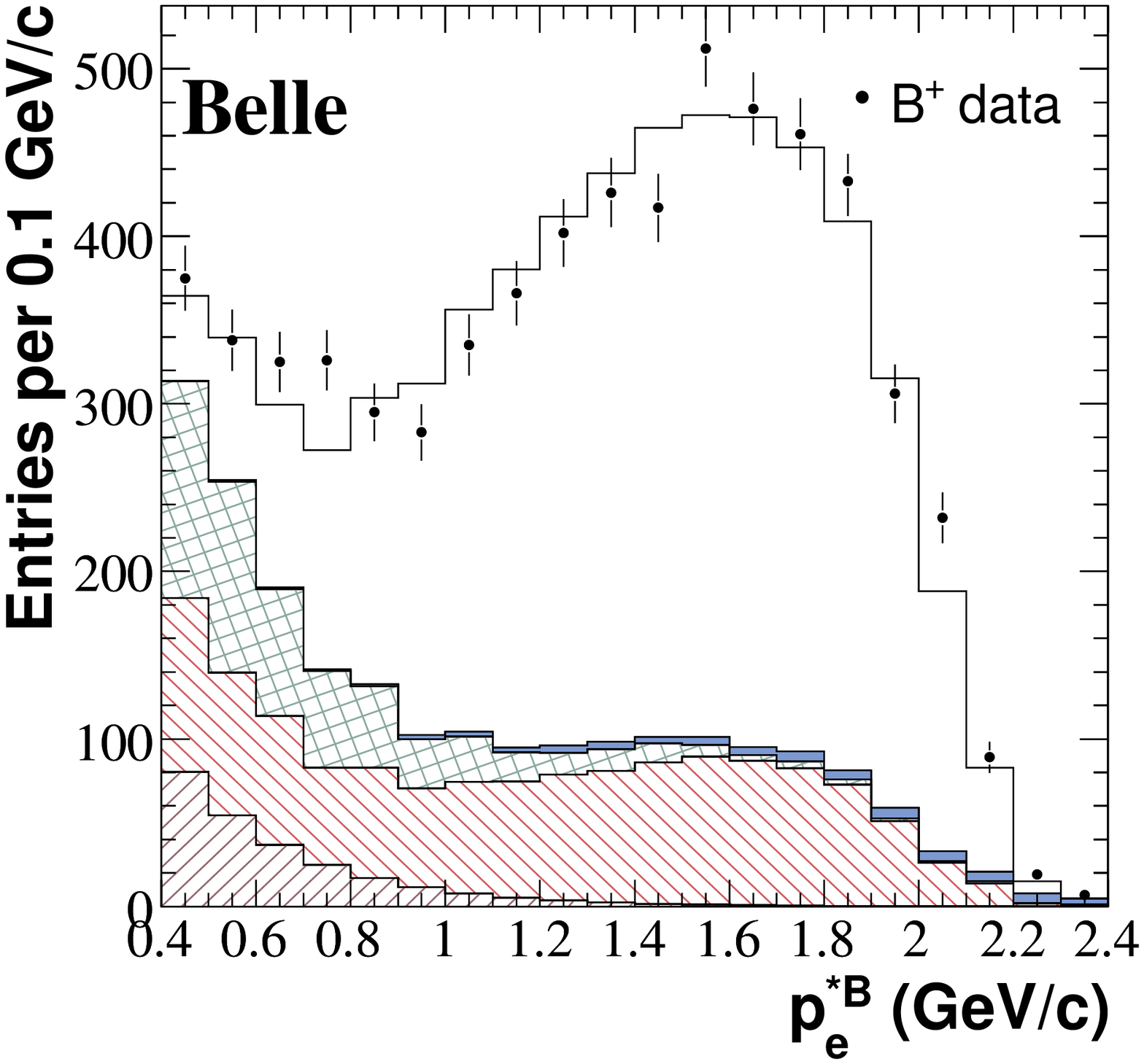}\\
    \includegraphics[width=0.9\columnwidth]{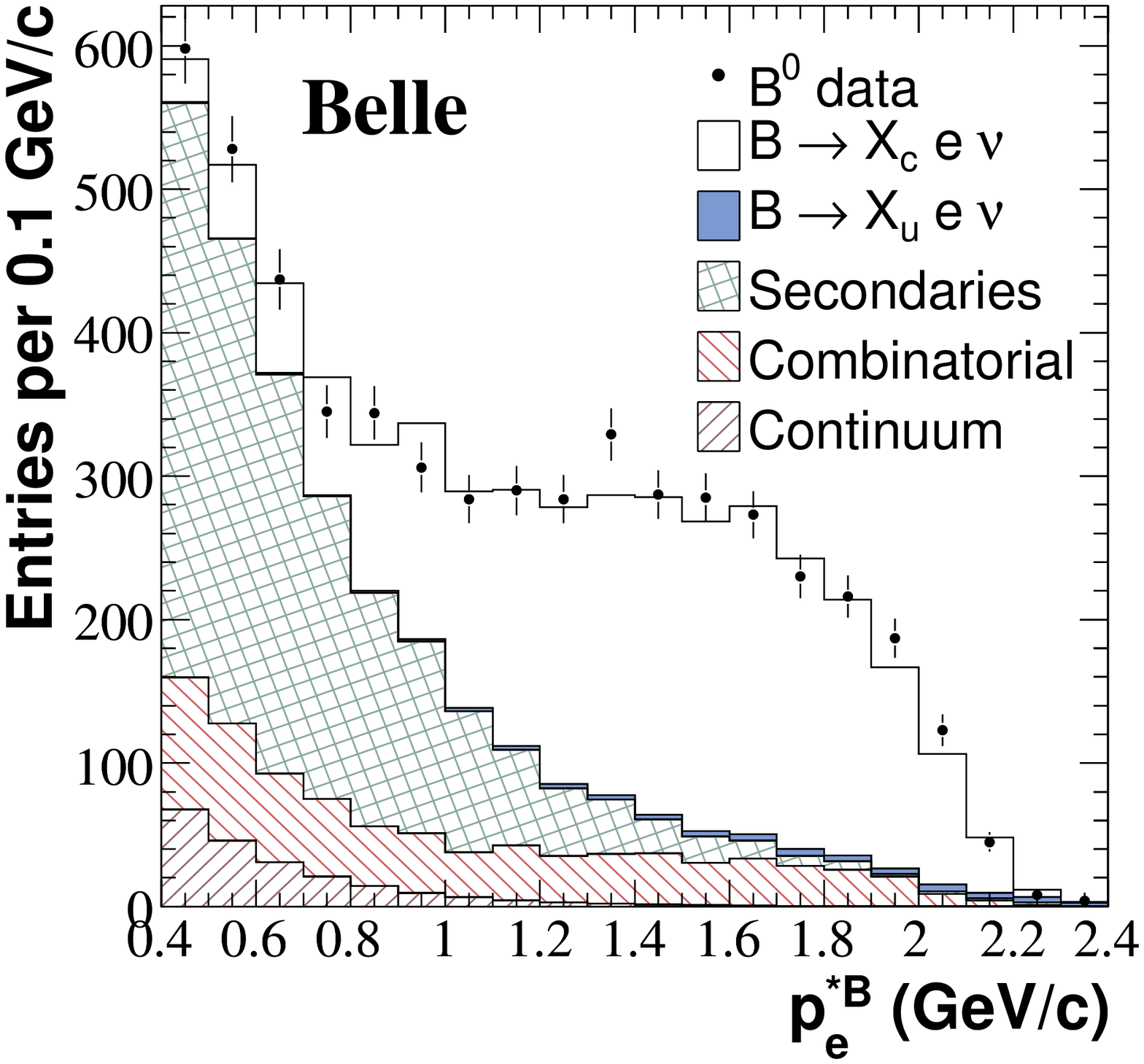}\\
  \end{center}
  \caption{Electron energy spectrum in the c.m.\ frame for charged
  (top) and neutral (bottom) tags, as measured in
  Ref.~\cite{Urquijo:2006wd}.} \label{fig:2}
\end{figure}

Belle has recently measured the c.m.\ electron energy~\cite{Urquijo:2006wd}
and the hadronic mass~\cite{Schwanda:2006nf} spectra in $B\to
X_c\ell\nu$~decays, based on 140~fb$^{-1}$ of $\Upsilon(4S)$ data. The
experimental procedure is very similar to the BaBar analysis, {\it
  i.e.}, one $B$~meson in the event is fully reconstructed in a hadronic
channel (Fig.~\ref{fig:2}). The main difference to the analysis
discussed above is that detector effects in the spectra are removed by
unfolding using the Singular Value Decomposition (SVD)
algorithm~\cite{Hocker:1995kb} with a detector response matrix found
from MC simulation. The moments are then calculated from the unfolded
spectra. In Ref.~\cite{Urquijo:2006wd}, Belle measures the partial
semileptonic branching fraction and the c.m.\ electron energy moments
$\langle E^n_e\rangle$, $n=1,\dots,4$, for minimum c.m.\ electron energies
ranging from 0.4 to 2.0~GeV. The hadronic mass
analysis~\cite{Schwanda:2006nf} measures the first and second moments
of $M^2_X$ for minimum c.m.\ lepton energies between 0.7 and 1.9~GeV.

An interesting analysis of inclusive $B\to X_c\ell\nu$~decays comes
also from the DELPHI experiment~\cite{Abdallah:2005cx}: In this study,
the $b$-frame lepton energy $\langle E^n_l\rangle$, $n=1,2,3$, and the
hadronic mass $M^{2n}_X$, $n=1,\dots,5$, moments are measured without
applying any selection on the lepton energy in the $b$-frame. This is
possible as DELPHI studies $b$-decays in $Z\to b\bar
b$~events. $b$-hadrons are thus produced with significant kinetic
energy in the laboratory frame, so that charged leptons at rest in the
$b$-frame can be seen in the detector.

\section{\boldmath Global fit and determination of $|V_{cb}|$
  inclusive}

The different global fit analyses differ by the theory expressions and
the data sets they are based onto: As discussed in Sect.~\ref{sect:2},
at present there are two sets of theoretical expressions available
for this analysis derived in the
kinetic~\cite{Benson:2003kp,Gambino:2004qm,Benson:2004sg} and 1S
schemes~\cite{Bauer:2004ve}. As for the data sets,
DELPHI~\cite{Abdallah:2005cx}, BaBar~\cite{Aubert:2007yaa} and
Belle~\cite{Schwanda:2008kw} have determined $|V_{cb}|$ inclusive from
their own data. We will discuss the recently published Belle analysis
in more detail. Finally, to achieve the ultimate statistical precision,
we will combine all available data to measure $|V_{cb}|$ and the
$b$-quark mass $m_b$.

In Ref.~\cite{Schwanda:2008kw}, Belle uses 25~measurements of the
lepton energy and hadronic mass moments in $B\to X_c\ell\nu$ and of
the photon energy moments in $B\to X_s\gamma$
(Table~\ref{tab:2}). Though more moment measurements are available,
moments without matching theoretical prediction or highly correlated
measurements are excluded. A $\chi^2$-fit of these measurements is
done to both the kinetic and 1S scheme expressions. The only
external input in the analysis is the average $B$~lifetime
$\tau_B=(1.585\pm 0.006)$~ps~\cite{Barberio:2007cr}.
\begin{table}
  \caption{Experimental inputs used in the Belle $|V_{cb}|$ inclusive
    analysis~\cite{Schwanda:2008kw}.} \label{tab:2}
  \begin{center}
    {\small \begin{tabular}{l|l}
      \hline \hline
      Moments & Measurements used\\
      \hline
      & $n=0$: $E_\mathrm{min}=0.6$, 1.0, 1.4~GeV\\
      Lepton energy & $n=1$: $E_\mathrm{min}=0.6$, 0.8, 1.0, 1.2,
      1.4~GeV\\
      $\langle E^n_\ell\rangle$~\cite{Urquijo:2006wd} & $n=2$:
      $E_\mathrm{min}=0.6$, 1.0, 1.4~GeV\\
      & $n=3$: $E_\mathrm{min}=0.8$, 1.0, 1.2~GeV\\
      \hline
      Hadronic mass & $n=1$: $E_\mathrm{min}=0.7$, 1.1, 1.3, 1.5~GeV\\
      $\langle M^{2n}_X\rangle$~\cite{Schwanda:2006nf} &
      $n=2$: $E_\mathrm{min}=0.7$, 0.9, 1.3~GeV\\
      \hline
      Photon energy & $n=1$: $E_\mathrm{min}=1.8$, 2.0~GeV\\
      $\langle E^n_\gamma\rangle$~\cite{Schwanda:2008kw} &
      $n=2$: $E_\mathrm{min}=1.8$, 2.0~GeV\\
      \hline \hline
    \end{tabular} }
  \end{center}
\end{table}

There a 7 free parameters in both fits. Only $|V_{cb}|$ can be
compared directly, while a scheme translation must be performed for
the HQ parameters, including the $b$-quark mass $m_b$. The main
challenge is to properly account for correlations: the
covariance matrix of the $\chi^2$-fit is the sum of experimental and
theoretical correlations. While the experimental correlation
coefficients have been determined in the respective
analyses~\cite{Urquijo:2006wd,Schwanda:2006nf,Schwanda:2008kw},
theoretical correlations are estimated following the prescriptions of
the theoretical
authors~\cite{Gambino:2004qm,Benson:2004sg,Bauer:2004ve}. The results
of the Belle analysis are given in Table~\ref{tab:3}.
\begin{table}
  \caption{Results of the Belle
    $|V_{cb}|$~analysis~\cite{Schwanda:2008kw}. The results for $m_b$ are
    compatible after scheme translation.} \label{tab:3}
  \begin{center}
    {\small \begin{tabular}{c|c|c}
      \hline \hline
      & Kinetic scheme & 1S scheme\\
      \hline
      $|V_{cb}|$ (10$^{-3}$) & $41.58\pm 0.69(fit)$ & $41.56\pm
      0.68(fit)$\\
      & $\pm 0.08(\tau_B)\pm 0.58(th)$ & $\pm 0.8(\tau_B)$\\
      \hline
      $m_b$ (GeV) & $4.543\pm 0.075$ & $4.723\pm 0.055$\\
      \hline
      $\chi^2$/ndf. & $7.3/18$ & $4.7/18$\\
      \hline \hline
    \end{tabular} }
  \end{center}
\end{table}

\begin{figure}
  \begin{center}
    \includegraphics[width=0.9\columnwidth]{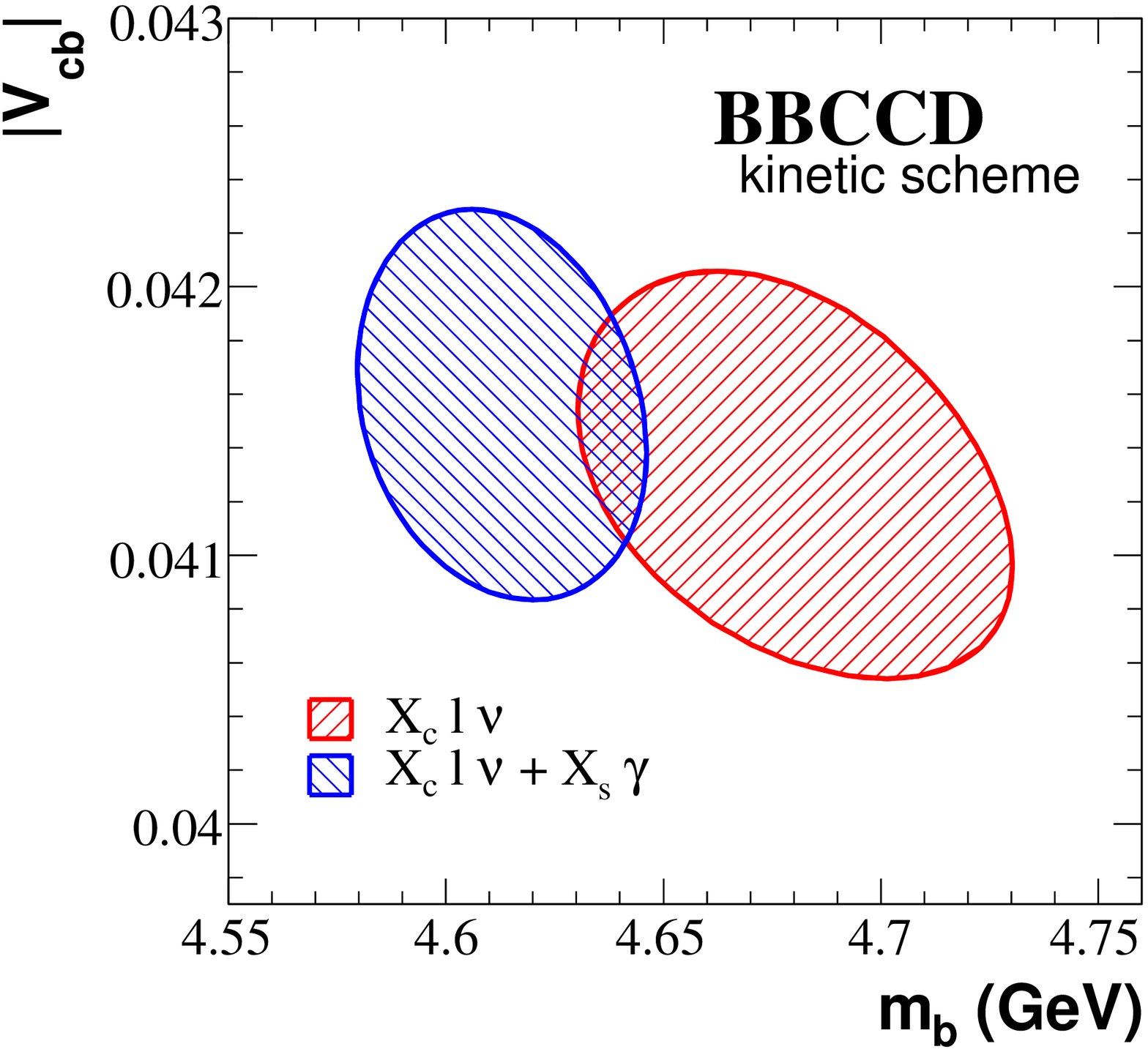}\\
    \includegraphics[width=0.9\columnwidth]{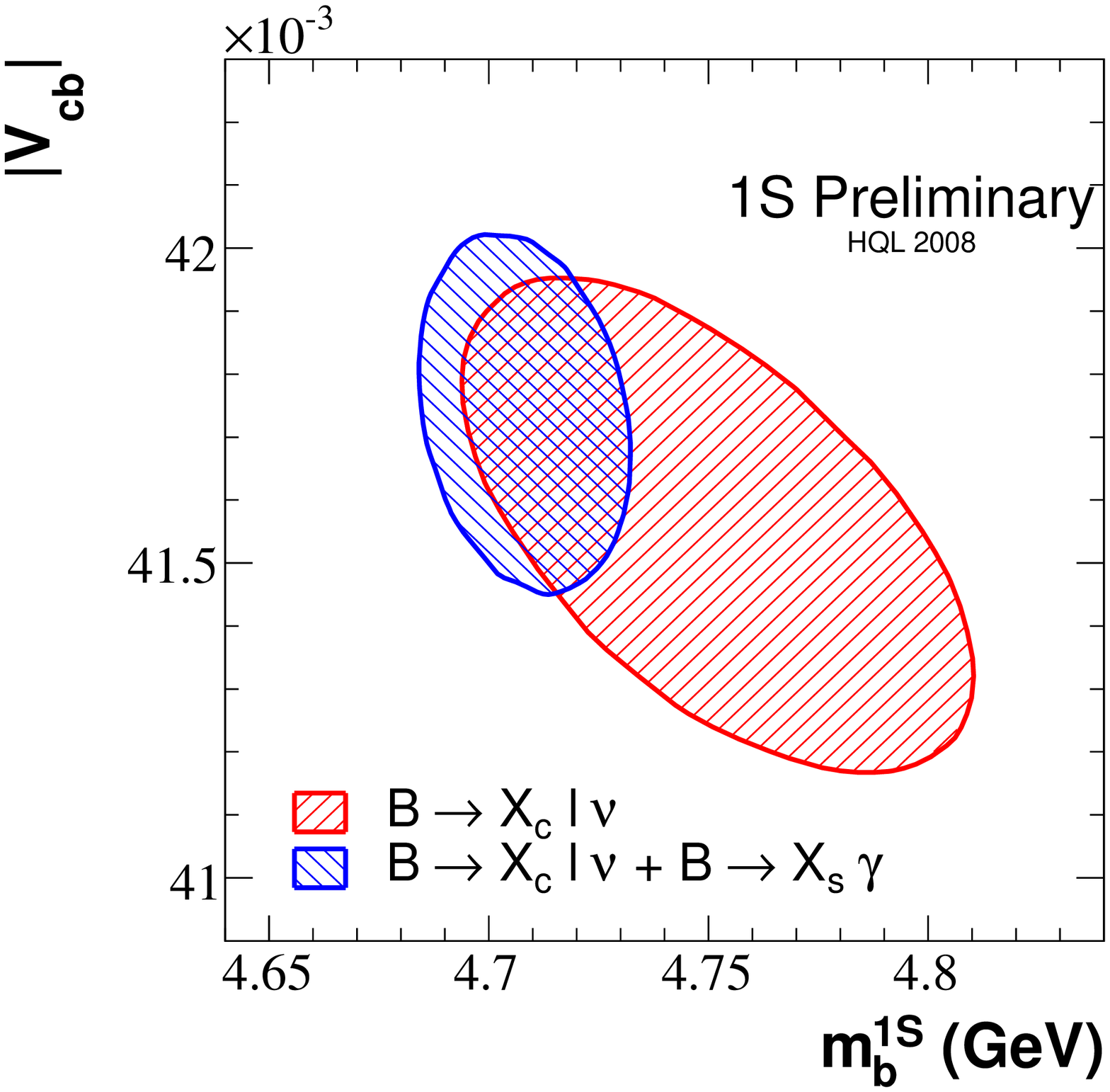}\\
  \end{center}
  \caption{Preliminary results of the fits in the kinetic (top) and 1S
    (bottom) schemes combining all available moment data
    (Table~\ref{tab:4}). The $\Delta\chi^2=1$~contours are shown for
    the fit with and without $B\to X_s\gamma$~data.} \label{fig:3}
\end{figure}

Finally, we attempt to combine all available moment measurements to
optimize the statistical precision in $|V_{cb}|$ and $m_b$. Using
70~measurements (Table~\ref{tab:4}) from different experiments, we
follow the Belle approach to derive numbers in the kinetic and 1S
schemes. The preliminary results are shown in Table~\ref{tab:5} and
Fig.~\ref{fig:3}.
\begin{table}
  \caption{Experimental inputs used for a global fit analysis to all
    available moment data.} \label{tab:4}
  \begin{center}
    {\small \begin{tabular}{l|l}
      \hline \hline
      Experiment & Measurements used\\
      \hline
      BaBar & $\langle E^n_\ell$, $n=0,1,2,3$~\cite{Aubert:2004td}\\
      & $\langle M^{2n}_X\rangle$, $n=1,2$~\cite{Aubert:2004tea}\\
      & $\langle E^n_\gamma\rangle$,
      $n=1,2$~\cite{Aubert:2006gg,Aubert:2005cua}\\
      \hline
      Belle & $\langle E^n_\ell$, $n=0,1,2,3$~\cite{Urquijo:2006wd}\\
      & $\langle M^{2n}_X\rangle$, $n=1,2$~\cite{Schwanda:2006nf}\\
      & $\langle E^n_\gamma\rangle$, $n=1,2$~\cite{Abe:2008sxa}\\
      \hline
      CDF & $\langle M^{2n}_X\rangle$, $n=1,2$~\cite{Acosta:2005qh}\\
      \hline
      CLEO & $\langle M^{2n}_X\rangle$, $n=1,2$~\cite{Csorna:2004kp}\\
      & $\langle E^n_\gamma\rangle$, $n=1$~\cite{Chen:2001fja}\\
      \hline
      DELPHI & $\langle E^n_\ell$, $n=1,2,3$~\cite{Abdallah:2005cx}\\
      & $\langle M^{2n}_X\rangle$, $n=1,2$~\cite{Abdallah:2005cx}\\
      \hline \hline
    \end{tabular} }
  \end{center}
\end{table}
\begin{table}
  \caption{Preliminary results of the analysis combining all available
    moment data (Table~\ref{tab:4}).} \label{tab:5}
  \begin{center}
    {\small \begin{tabular}{c|c|c}
      \hline \hline
      & Kinetic scheme & 1S scheme\\
      \hline
      $|V_{cb}|$ (10$^{-3}$) & $41.55\pm 0.43(fit)$ & $41.74\pm
      0.29(fit)$\\
      & $\pm 0.08(\tau_B)\pm 0.58(th)$ & $\pm 0.8(\tau_B)$\\
      \hline
      $m_b$ (GeV) & $4.613\pm 0.033$ & $4.708\pm 0.024$\\
      \hline
      $\chi^2$/ndf. & $30.6/63$ & $26.1/63$\\
      \hline \hline
    \end{tabular} }
  \end{center}
\end{table}

\section{Summary}

We have reviewed the theory and surveyed the experimental data for the
determination of the CKM matrix element $|V_{cb}|$ from inclusive
decays $B\to X_c\ell\nu$. The results for $|V_{cb}|$ using the data of
the Belle experiment alone are given in Table~\ref{tab:3}. Also, an
attempt is made to combine all available experimental data: The
preliminary results in terms of $|V_{cb}|$ and the $b$-quark mass
$m_b$ are shown in Table~\ref{tab:5}.

\section*{Acknowledgments}
% Please paste this acknowledgement into your latex file. 
% updated 12/15/08   add Nagoya's TLPRC, 2 Grant-in-Aids (long only)
%                        2 new NNSFC contract no. (long only) 
% updated 11/26/08   Poland: KBN -> MNiSW, Australia: DEST -> DISR
%
%***** Acknowledgments *****
%----------- Long version, for most papers ----------- 
We thank the KEKB group for the excellent operation of the
accelerator, the KEK cryogenics group for the efficient
operation of the solenoid, and the KEK computer group and
the National Institute of Informatics for valuable computing
and SINET3 network support.  We acknowledge support from
the Ministry of Education, Culture, Sports, Science, and
Technology (MEXT) of Japan, the Japan Society for the 
Promotion of Science (JSPS), and the Tau-Lepton Physics 
Research Center of Nagoya University; 
the Australian Research Council and the Australian 
Department of Industry, Innovation, Science and Research;
the National Natural Science Foundation of China under
contract No.~10575109, 10775142, 10875115 and 10825524; 
the Department of Science and Technology of India; 
the BK21 program of the Ministry of Education of Korea, 
the CHEP src program and Basic Research program (grant 
No. R01-2008-000-10477-0) of the 
Korea Science and Engineering Foundation;
the Polish Ministry of Science and Higher Education;
the Ministry of Education and Science of the Russian
Federation and the Russian Federal Agency for Atomic Energy;
the Slovenian Research Agency;  the Swiss
National Science Foundation; the National Science Council
and the Ministry of Education of Taiwan; and the U.S.\
Department of Energy.
This work is supported by a Grant-in-Aid from MEXT for 
Science Research in a Priority Area (``New Development of 
Flavor Physics''), and from JSPS for Creative Scientific 
Research (``Evolution of Tau-lepton Physics'').

\end{document}